# Five-fold symmetry in the hydrogen atom probed with accurate 1S-nS terms

G. Van Hooydonk, Ghent University, Faculty of Sciences, Ghent, Belgium

Abstract. We probe Penrose's five-fold symmetry in the hydrogen atom using its radius $r_H$ as derived classically from its mass $m_H$. This generic H symmetry, obeying Euclid's golden ratio [sqrt(5)-1]/2 for its 2 constituent complementary parts, is confirmed with precise H terms. These give away a Hund-type Mexican hat curve for natural H, which points to its mirrored variant, antihydrogen $\underline{H}$. We predict that term H 1S-3S, to be measured soon, is 2 922 743 278 654(2) kHz.

## I. Introduction

Uncertainties with H CPT-symmetry resulted in attempts to produce antihydrogen $\underline{H}$, to measure its interval $\underline{H}$ 1S-2S and to compare this with H 1S-2S [1]. Instead of CPT, Euclidean symmetries also rationalize composite unit H. Euclidean numbers appear in many fields of science [2-3], for chaotic/fractal behavior (Mandelbrot [4], Gutzwiller [5]) and for Penrose 5-fold symmetry [6]. Whenever parts $+x$ and $(1-x)$ in unit 1 obey $(1-x)/x=x/1$, Euclidean symmetries $x_\pm=\varphi_\pm=\tfrac{1}{2}(1\pm\sqrt{5})$ appear. Atom H, the simplest composite but most abundant unit in the Universe [7], has complementary parts electron $(m_e)$ and proton $(m_P)$: $m_H=m_e+m_p=m_e+(m_H-m_e)$ or $1=x+(1-x)$, if $x=m_e/m_H$. If $\varphi$ were relevant for H, it must show in its spectrum, although it is absent in QED [8]. We now probe $\varphi$ for H using mass $m_H$ and radius $r_H$ in $m_H=4\pi\gamma r_H^3/3$. In line with Rydberg's formula [9], scaling H levels by $\tfrac{1}{2}e^2/r_H$ gives away $\varphi$ and fractal behavior with precise H 1S-nS terms [10]. We predict a value of 2922743278654(2) kHz for H 1S-3S, to be measured soon [11].

## II. Rydberg equation and fractal behavior of atom H

*II.1 Chaotic/fractal interpretation of the Rydberg formula for composite H*

With constant a in Å and line number n, the original Rydberg formula [9] for H terms

$$T_n = an^2/(n^2-1) \text{ Å or } T_n/(an) = n/(n^2-1) = 1/(n-1/n) \qquad (1)$$

suggests that H exhibits fractal/chaotic behavior [4,5]. Bohr energy differences

$$\Delta E_n = 1/T_n = (n^2-1)10^8/(an^2) = R_H(1-1/n^2) = R_H - R_H/n^2 = E_n - E_1 \text{ cm}^{-1} \qquad (2)$$

with Rydberg $R_H=10^8/a$ cm$^{-1}$, give the linear version of (1), i.e.

$$n\Delta E_n/R_H = (n-1)(n+1)/n = n-1/n \qquad (3)$$

With $E_n$ [12] instead of $\Delta E_n$, plots of $nE_n$ versus n and 1/n give power laws

$$E_n(n) \equiv E_n(1/n) = 109679{,}223605211\, n^{-1{,}000004252339} \equiv 109679{,}223605211(1/n)^{1{,}000004252339} \qquad (4)$$

Linear n and inverse 1/n views on fractal H (1)-(3) give errors of only 0,007 cm$^{-1}$, while Bohr's are 0,0126 cm$^{-1}$ (a power fit in $1/n^2$ has its exponent shifted by 1). Fractal asymptote 109679,2236 cm$^{-1}$ in (4) is much larger than in Bohr theory or QED, i.e. $-E_1$=109678,773704 cm$^{-1}$ [12], to which we return in Section V. Since 1/n secures convergence, a 4$^{th}$ order fit in 1/n

$$nE_n = 0{,}006889343262/n^4 - 4{,}375765800476/n^3 + 5{,}5580713748932/n^2 + 109677{,}585385323000/n \qquad (5)$$

is accurate within $10^{-8}$ cm$^{-1}$ or 0,45 kHz (less precise data [13] behave similarly). By its precision, (5) must be important for H-based metrology [10, 14-15].

*II.2 Generalized Bohr H theory and reduced mass: opening for φ*

To not to interrupt the argument on φ, we compare H theories in Appendix A. With (A1)-(A2), Bohr's integer quantum number n and Rydberg $R_H$ give rotational level energies

$$E_n = -R_H/n^2 = -\tfrac{1}{2}(\hbar^2/\mu e^2)/n^2 = -\tfrac{1}{2}\mu\alpha^2 c^2/n^2 = -\tfrac{1}{2}(e^2/r_0)/n^2 \qquad (6)$$

Here, $r_0$ is Bohr radius $r_B = \hbar^2/(m_e e^2)$, corrected for reduced electron mass μ, according to

$$\mu = m_e m_P/(m_e + m_P) = m_e m_P/m_H = m_e/(1 + m_e/m_P) \equiv m_e(1 - m_e/m_H) \qquad (7)$$

Generalizing (6) with a critical n-value $n_c$ for another H radius $r_H$ gives respectively

$$r_H = n_c^2 r_0 \qquad (8)$$

$$E_n = -(R_H/n_c^2)(n_c/n)^2 = -\tfrac{1}{2}(e^2/r_H)(n_c/n)^2 \qquad (9)$$

(i) Any critical $n_c$ ($\neq 0$) will lead to the same accuracy as (6). A relation between $n_c$ and φ like

$$n_c = A\varphi^m \qquad (10)$$

plugged in (9), may probe Euclidean symmetry but only if an alternative $r_H$ really existed (see Section II.3). If not, both (8) and (9) are trivial.

(ii) Detecting internal φ-effects in H depends on specific φ-relations [2-3] like

$$\varphi^{m+2} + \varphi^{m+1} = \varphi^m; \quad 1 = 1/\varphi - \varphi; \quad \varphi^2 + \varphi - 1 = 0 \text{ and } \varphi(\varphi+1) = 1 \qquad (11)$$

Internal φ-symmetries (11) are available from dimensionless (7). Scaling by $m_H$ gives product

$$\varrho_H = \mu/m_H = (m_e/m_H)(1 - m_e/m_H) = x(1-x) \qquad (12)$$

for parts. Here, $d\varrho/dx = 1 - 2x = 0$ gives $\varrho_{max} = \tfrac{1}{4}$ for equal parts $x = \tfrac{1}{2}$, while $\varrho = x(1-x)$ or $x^2 - x + \varrho = 0$ gives $x_\pm = \tfrac{1}{2}[1 \pm \sqrt{(1-4\varrho)}]$. A center between parts gives $-x$, $(1-x)$, $x^2 - x + \varrho = 0$ and $x_\pm = \tfrac{1}{2}[1 \pm \sqrt{(1+4\varrho)}]$. Part ratios (12) secure all symmetries in (ii) are Euclidean. With $\varrho_{max} = \tfrac{1}{4}$ and (10), we further get

$$\varrho_H = x(1-x) \sim A\varphi^m(1 - A\varphi^m) \text{ and } \Delta\varrho = \varrho_{max} - \varrho = \tfrac{1}{4} - x(1-x) = \tfrac{1}{4}(1-2x)^2 \sim \tfrac{1}{4}(1-2A\varphi^m)^2 \qquad (13)$$

i.e. a parabola for φ-symmetries, with small corrections to $E_n$, since $\mu/m_e$ in (7) is $1837(\mu/m_H)$. The fate of parabolic φ-symmetry (13) for H depends on the reality of a valid alternative radius $r_H$.

*II.3 Alternative classical H radius $r_H$*

Apart from [16], a first principles alternative quantum radius for H, other than Bohr length $r_B$, does not exist. Only a classical 19th century macroscopic view on spherical H can give $r_H$ using

$$m_H = (4\pi/3)\gamma r_H^3 \text{ and } r_H = [(3/4\pi\gamma)m_H]^{1/3} \qquad (14)$$

with form factor $4\pi/3$ γ in g/cm$^3$ is H density, which fixes the external H symmetry (its form). With $m_H = m_e + m_P = 9,10938215 \cdot 10^{-28} + 1,672621637 \cdot 10^{-24}$ g [10] and γ=1 g/cm$^3$ for H, the result is

$$r_H = 7,36515437 \cdot 10^{-9} \text{ cm} = 0,736515437 \text{ Å} \qquad (15)$$

This is the only real, theoretically plausible alternative to Bohr length $r_B = 0,529177209$ Å [16].

In (6), H radius $r_0$ is Bohr length $r_B$, corrected for recoil (7) or

$$r_0 = [\hbar^2/(m_e e^2)](1 + m_e/m_p) = 0,5294654075 \text{ Å} \quad (16)$$

The ratio of classical natural H radius $r_H$ in (15) and Bohr's $r_0$ in (16) is

$$x = r_H/r_0 = 1,391054876\ldots \quad (17)$$

(without recoil, $r_H/r_B = 1,391812469\ldots$).

The natural virial Coulomb energy $-½e^2/r_H$ for any two charge-conjugated parts amounts to

$$½e^2/r_H = 78844,900590508 \text{ cm}^{-1} = 2363710654879,4 \text{ kHz} \quad (18)$$

### III. Scaling $E_n$ by $½e^2/r_H$: probing five-fold or $\varphi$-symmetry in atom H

Scaling $E_n$ by natural H asymptote (18) gives numbers

$$N_n = [E_n/(½e^2/r_H)]/n^2 \text{ or } nN_n = [E_n/(½e^2/r_H)]/n \quad (19)$$

Due to (18), plots of $nN_n$ versus $1/n$ and $(1-1/n)$ in Fig. 1 give 4$^{th}$ order fits (with 5 decimals)

$$N_n = -0,00006/n^4 + 0,00007/n^3 + 1,39106/n^2 \quad (20)$$

$$N_n = -0,000056(1-1/n)^4 + 0,00015(1-1/n)^3 + 1,39093(1-1/n)^2 - 2,78210(1-1/n) + 1,39107 \quad (21)$$

With $(1-1/n)$, typical for molecular potentials [16], (20)-(21) reveal the effect of odd powers in $1/n$, absent in Bohr $1/n^2$ theory and in a relativistic expansion in $E_n = \mu c^2[1/\sqrt{(1+\alpha^2/n^2)} - 1]$ [8,14].

In (A16)-(A17), we prove that the H *force constant* $k_n$, away from critical configuration $n_c$, varies with $1,5/n$. Fig. 1 includes $N_n$ versus $1,5/n$ and $(1-1,5/n)$ with 5-decimal 4$^{th}$ order fits

$$N_n = -0,00001(1,5/n)^4 + 0,00002(1,5/n)^3 + \mathit{0,61825}(1,5/n)^2 \quad (22a)$$

$$N_n = -0,00001(1-1,5/n)^4 + 0,00002(1-1,(/n)^3 + \mathit{0,61824}(1-1,5/n)^2 - \mathit{1,23651}(1-1,5/n) + \mathit{0,61826} \quad (22b)$$

Coefficients of $(1,5/n)^2$ in (22a) and $(1-1,5/n)^2$ in (22b) are close to Euclid or Phidias number (10)

$$\varphi = ½(\sqrt{5} - 1) = 1/\varphi - 1 = \Phi - 1 = 0,618034 \ldots \quad (23)$$

Correction factor $f_\varphi$ for external $\varphi$-symmetry and $f_r$ for recoil (for an internal H-symmetry)

$$f_\varphi = 0,618247/0,618034 - 1 = 0,000344; \quad f_r = m_e/m_p = 1/1836,15267247 = 0,000545 \quad (24)$$

shows that $f_\varphi$ is smaller than $f_r$ by 40 %. Difference $\delta$ for $\varphi$-symmetry is 0,02 %, i.e.

$$\delta = 0,618247 - 0,618034 = 0,000213 \quad (25)$$

In terms of ratio $m_e/m_H = 1/1837,15267247$ in (7), difference (25)

$$(m_H/m_e) 0,000213 = 0,390635 \approx (9\varphi/4 - 1) = (9/4)(½\sqrt{5} - 17/18) \quad (26)$$

reflects the importance of Euclid's golden ratio for H.

Combining coefficient for $1,5/n$ (22a) and asymptote 0,618247 in (22a-b) gives ratio x in (18), since

$$x = (9/4) \cdot 0,618246619 = (3/2)^2 \varphi = 1,391054894 = r_H/r_0 \quad (27a)$$

Using (9), the Euclidean H variable $x_E$ must therefore obey

$$x_E = a\varphi^{½}/n \quad (27b)$$

Results (21)-(27) probe Penrose's five-fold or Euclid's $\varphi$-symmetry in H, due to alternative classical radius $r_H$ (15).

If external symmetry (27) also applied for internal H-symmetry according to (13), (27) prescribe Euclidean variable and symmetry parabola, given respectively by

$$X_E \sim x_E(1-x_E) \sim (\varphi^{1/2}/n)(\varphi^{1/2}/n-1) \text{ and } \Delta\varrho = \varrho_{max} - \varrho \sim \tfrac{1}{4}(1-2\varphi^{1/2}/n)^2 \qquad (28)$$

whenever a=1 in (27b). As shown previously [22], Euclidean symmetry parabola (28) appears in higher order for the H Lyman series 1s-nS [12]. The phenomenological H parabola $(1-1,572273/n)^2$ in [22] validates theoretical Euclidean H parabola (28), since $2\varphi^{1/2}=1,572303$. The difference with 1,572273 in [22] is only 0,002 %. We now verify whether precise H terms [17-21] confirm this early evidence [22] for 5-fold H symmetry (28).

**IV. Accurate H intervals (prediction of H $1S_{1/2}$-$3S_{1/2}$)**

The precision needed to validate (28) requires an upgrade of $E_n$ [12], used in [22]. Table 1 shows the H terms available: 4 precisely known terms A, B, D and E give 2 derived terms C and F. Since only B and F are void of 1S, the immeasurable series limit or $-E_1$, B and F allow a simple conversion. Precision at this level requires many significant digits. A fit of $E_n$ [12] to 4$^{th}$ order in 1/n through the origin, tested with terms in Table 1, gives slope $1-1,79201817 \cdot 10^{-8}$ and intercept $26940,95752/29979245,8=0,00008965361$ cm$^{-1}$. This results in the terms in Table 2. The conversion corresponds with a change of Erickson's 1977 Rydberg $R=109737,3177\pm 0,00083$ cm$^{-1}$ [12].

Table 1 reveals that A, B and C are exactly reproduced. The small discrepancies for D, E and F are much lower than experimental uncertainties, 10 kHz for D and 21 for E in [20-21]. With the small error of 1,74 kHz for F removed, the error reappears for D and E (1,71 kHz). The small difference of 1,26 kHz for all terms caused by this correction justifies their omission in Table 2.

Table 1 Observed [10] and intervals from this work in kHz (with errors δ). Prediction of H 1S-3S

| Intervals[a,b] | Observed | This work | δ(kHz) | Ref[c]. |
|---|---|---|---|---|
| A. 1S-2S | 2466061413187,07 | 2466061413187,07 | 0,00 | [17,18] |
| B. 2S-8S | 770649350012,00 | 770649350012,00 | 0,00 | [19] |
| C. [1S-8S] | 3236710763199,07 | 3236710763199,07 | 0,00 | |
| D. 2S-4S-¼(1S-2S) | 4797338 | 4797334,20 | -3,80 | [20] |
| E. 2S-6S-¼(1S-3S) | 4197604 | 4197601,94 | -2,06 | [21] |
| F. [6S-2S+¼(3S-2S)] | 599734 | 599732,26 | -1,74 | |
| G. **1S-3S** predicted[d] | to be measured | **2922743278654,37** | | [11] |

[a] only B and derived F do not depend on 1S
[b] derived values between square brackets result from C=A+B and F=D-E
[c] the four intervals A,B,D,E are used for metrology in [10]
[d] by the same argument, all other intervals nS in Table 1 are predicted with the relative accuracy to reference term B [19]

Referring to [11], the predicted H 1S-3S interval (G in Table 1, n=3 in Table 2) is correct within 1,74 kHz, i.e. the largest error in Table 2.

A 4th order fir is still sufficient to fit all data accurately, when 15 significant digits are used.

$N_n = E'_n/(\frac{1}{2}e^2/r_H)$ plotted versus Euclidean variable $x_E$ (27b) gives $N_n$, equal to

$-0,000028651871617 x_E^4 + 0,000042968542402 x_E^3 + 1,000344034289810 x_E^2 - 0,000000000165642 x_E$ (29)

For 19 terms 2S to 20S in Table 2, average errors of 0,11 kHz give a precision of $1,6 \cdot 10^{-12}$ %. Small deviations $\varepsilon_n$ nevertheless increase with increasing n (which we discuss elsewhere).

H terms in Table 2 allow a check of Euclidean variable $X_E$ (28) for internal Euclidean $\varphi$-symmetry.

Table 2 H 1S-nS: original $E_n$ [12] and converted $E'_n$ in cm$^{-1}$, terms $T_n$ in kHz and deviations $\varepsilon_n$ with fitting to 4th order (29)

| n | $-E_n$ (cm$^{-1}$) | $-E'_n$ (cm$^{-1}$) | $T_n$ (kHz) | $\varepsilon_n$ (kHz) |
|---|---|---|---|---|
| 1 | 109678,773704000 | 109678,77174307900 | 0 | |
| 2 | 27419,817835200 | 27419,81734379700 | 2466061413187,07 | 1,706 |
| 3 | 12186,550237200 | 12186,55001899660 | 2922743278654,37 | 0,139 |
| 4 | 6854,918845390 | 6854,91872213227 | 3082581563818,04 | -0,078 |
| 5 | 4387,140880900 | 4387,14080222353 | 3156563684658,80 | -0,097 |
| 6 | 3046,621950400 | 3046,62189584705 | 3196751430452,60 | -0,083 |
| 7 | 2238,332451300 | 2238,33241135261 | 3220983339585,82 | -0,065 |
| 8 | 1713,722059150 | 1713,72202861737 | 3236710763199,07 | -0,050 |
| 9 | 1354,051221430 | 1354,05119731790 | 3247493423457,69 | -0,038 |
| 10 | 1096,780974420 | 1096,78095487230 | 3255206191292,99 | -0,029 |
| 11 | 906,430202530 | 906,43018635921 | 3260912763770,46 | -0,022 |
| 12 | 761,652903990 | 761,65289037408 | 3265253077913,06 | -0,016 |
| 13 | 648,982171840 | 648,98216020327 | 3268630861427,32 | -0,012 |
| 14 | 559,581428918 | 559,58141885409 | 3271311028226,93 | -0,008 |
| 15 | 487,457495457 | 487,45748665884 | 3273473249318,27 | -0,005 |
| 16 | 428,429358101 | 428,42935033704 | 3275242868326,18 | -0,003 |
| 17 | 379,508294780 | 379,50828787203 | 3276709484882,61 | -0,001 |
| 18 | 338,511977355 | 338,51197116509 | 3277938523538,06 | 0,000 |
| 19 | 303,816802757 | 303,81679717463 | 3278978658687,20 | 0,001 |
| 20 | 274,194630876 | 274,19462581233 | 3279866709043,60 | 0,002 |
| | | | average | 0,124 |

## V. Probing internal $\varphi$-symmetry for fractal H

A 4th order fit of accurate $E'_n$ data in Table 2 exposes the contribution of Bohr's $1/n^2$ theory

$-E'_n = -4,368336200714/n^4 + 5,555412530899/n^3 + 109677,583783388/n^2 - 0,000015348196/n$ (30)

Subtracting term $1/n^2$ discloses accurately symmetry differences beyond Bohr theory

$\Delta E'_n = (4,368336200714/n^2 - 5,555412530899/n)/n^2$ cm$^{-1}$ (31)

if the small $1/n$ term is disregarded. Using $E_1$ in Table 2 gives $\Delta E'_n$, shifted by $1,18.../n^2$ (see [22] for the analysis based on $E_n$). The hidden parabola in (31) is obtained by adding $(\frac{1}{2}5,5554/\sqrt{4,3683})^2$ $=1,32901^2 = 1,766268$. This leads to a harmonic Rydberg $R_{harm}$, larger than $R_\infty$ and $R_1$ and as revealed by power fit (4) above, equal to [22]

$R_{harm} = 109677,583783 + 1,766268 = 109679,350051$ cm$^{-1}$ (32)

H symmetry equation (31) with $R_{harm}$ becomes a perfect Mexican hat curve, i.e. quartic [23]

$$\Delta_{harm}=(4{,}368336/n^2-5{,}555413/n+1{,}766268)/n^2 \text{ cm}^{-1}=1{,}766268(1-1{,}572642/n)^2/n^2 \quad (33)$$

critical at n=2.1,572642/n≈π≈4φ½ [23].

Fig. 2 illustrates quartics for $R_{harm}$, $R_\infty$ and $E_1$ versus 4φ½/n-1. The Hund-type Mexican hat curve with $R_{harm}$ (33) is a signature for left-right asymmetry for composite atom H [23], most likely also the missing link to rationalize in- and external H symmetries, including H [22,23]. Using $R_\infty$ to disclose internal H symmetries as in QED creates larger energy differences (see Fig. 2). With (33) accurate to order kHz, five-fold H symmetry is obvious, since the theoretical Euclidean symmetry parabola (28) is reproduced exactly by the experimental H data in Table 2.

In fact, all numbers in (33) are sufficiently close to Euclidean variables (27)-(28), i.e.

$$9φ^{½}/4=1{,}768840600 \quad (34)$$

$$2φ^{½}=1{,}572302756 \quad (35)$$

transforming (33) in $(9φ^{½}/4)(1-2φ½/n)^2/n^2$ and (31) in $(9φ^{½}/4)[1-(1-2φ½/n)^2]/n^2$. For internal H symmetry (35), the difference is only 0,000338763, just like 0,000344 in (24). This proves that both in- and external H symmetries stem from chaotic/fractal behavior [4-5], Euclid's golden number [2-3] or Penrose's 5-fold symmetry [6], the most important, almost divine symmetry in nature [2-3].

**VI. Discussion**

(i) Spectral H data accurately follow a closed form quartic in 1/n. Unless for Lamb shifts, odd 1/n powers are absent in Bohr $1/n^2$ and QED theories. If observed data [13] had 5 decimals, the use of [12] could have been avoided, since all main intervals in Table 1 are assessable with [13]. Only the smaller intervals remain with an error using [13] for conversion (for F in Table 1, this error of 100 kHz suggests Kelly data [13] have a wrong 4th decimal for 4S and/or 6S).

(ii) Euclidean H harmony rests on algebra, overlooked for H parts, e.g. recoil [16], see Section II.2. Like Cagnac et al. [14], we find that that using μ as in relativistic theory (Section III) is questionable.

(iii) In the $H_2$ spectrum, natural asymptote $½e^2/r_H≈78844{,}9$ cm$^{-1}$ shows as ionic energy $D_{ion}= e^2/r_H$ [16]: $r_H$ is close to observed separation 0,74 Å in $H_2$ [24] and gives fundamental $H_2$ frequency of 4410 cm$^{-1}$ [24]. With $r_H$ and φ, molecular $H_2$ and atomic H spectra are intimately linked [16].

(iv) Incidentally, an angle of 30°, typical for Euclid's φ, also appears in the SM [25] as mixing angle for perpendicular interactions.

(v) Higher order terms in ξ=a/n or (1-ξ) brings H theory in line with Kratzer-type expansions like

$$E_n=a_0ξ^2 (1+ a_1ξ + a_2ξ^2 + a_3ξ^3+…) \quad (36)$$

formally similar to but different than the more familiar Dunham expansion [26-30].

(vi) With the Sommerfeld-Dirac fine structure formula [31], the internal variable for H nP is 1,5/n, rather than (35) for nS, which is responsible for the observed Lamb shifts [22, 31].

Euclidean H-symmetry, brought in by natural radius $r_H$ is in line with the original Rydberg equation (1) and connects all H terms directly with its most important property, mass $m_H$. H is prototypical for atomic and molecular physics [16] as well as for fractal behavior, in line with Mandelbrot's views [4]. We do not expand on H structures, conforming to 5-fold symmetry [6], since Bohr's model is likely to be revised with classical physics, following the lines set out in Appendix A.

**VII. Conclusion**

Although overlooked for a century, Euclidean H symmetry only shows when H mass is linked to its spectrum by its natural, classical radius $r_H$. A Hund-type Mexican hat curve for natural H proves that H is left-right asymmetric, which points to left- and right-handed states for composite hydrogen, say H and H̲ [22,23,29]. While this diverges from CPT-views on H [1], validating Euclidean φ-views on H rests, for a large part, on the value of H 1S-3S, to be measured in the near future [11].

Fig. 1 $nN_n$ versus $1/n$ (Δ), $1-1/n$ (□) (solid lines), $1,5/n$ (+) and $1-1,5/n$ (x) (dashes).

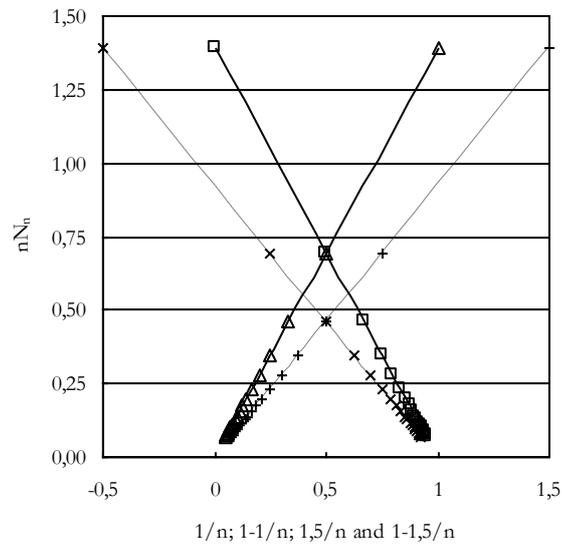

Fig. 2 Symmetry breaking curves in Euclidean H: $E_n$-differences (31)-(33) in cm$^{-1}$ versus the appropriate Euclidean variable, see text): Mexican hat curve with $R_{harm}$ (full-line □), with $E_1$ (short dashes o) and with NIST's $R_\infty$ (broken dashes +)

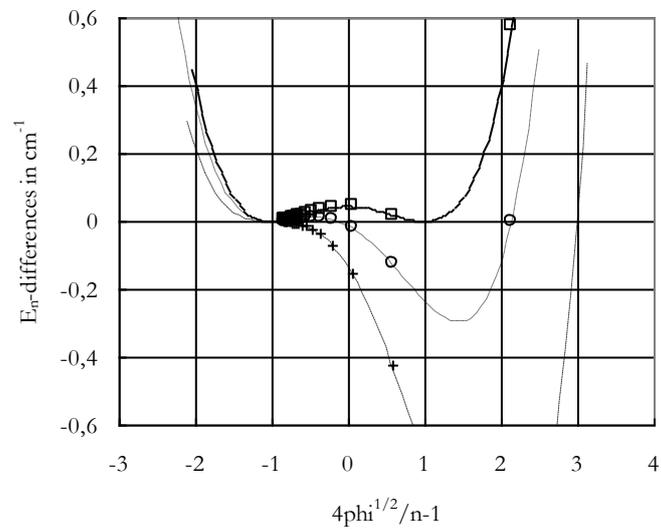

Appendix A Comparison of classical and Bohr H theories

This self-explanatory table contains all formulae for a stable charge-conjugated two particle Coulomb system, subject to periodic motion. Main results and differences are in bold.

| Description | Classical H theory | Bohr H theory | # |
|---|---|---|---|
| Energy E=T+V | $E=\frac{1}{2}\mu v^2-e^2/r$ | idem | A1 |
| Hamiltonian | $E=\frac{1}{2}p^2/\mu-e^2/r$ | idem | A2 |
| Periodic motion | $E=\frac{1}{2}\mu\omega^2 r^2-e^2/r$ | idem | A3 |
| Repulsive force d/dr | $\mu\omega^2 r=\mu v^2/r=p^2/(\mu r)$ | idem | A4 |
| Attractive force d/dr | $e^2/r^2$ | idem | A5 |
| Equal forces (Newton) | $\mu v^2 r=e^2$ | idem | A6 |
| Equal forces (Keppler, HO[a]) | $\mu v^2=e^2/r$; $\mu\omega^2=e^2/r^3$; $\omega^2=e^2/\mu r^3$; $\omega=\sqrt{(k/\mu)}$ | vibrator or HO not considered | A7 |
| Force constant $k_e$ at $r_e$ | $k_e=e^2/r_e^3$ | absent | A8 |
| Constant periodicity dE/dω | $\mu\omega r^2=\mu vr=pr=C$ | $\mu\omega r^2=\mu vr=pr=n\hbar$ | A9 |
| Moment | $p=C/r$ | $p=n\hbar/r$ | A10 |
| Ratio A6/A9 | $v=e^2/C$ | $v=e^2/(n\hbar)$; $v/c=e^2/(n\hbar c)=\alpha/n$ | A11 |
| H radius | $r=C/(\mu v)=C^2/(\mu e^2)$ | $r=n\hbar/(\mu v)= n^2\hbar^2/(\mu e^2)=n^2 r_B$ | A12 |
| Feedback of A10 in E (A1) | $\frac{1}{2}p^2/\mu-e^2/r=\frac{1}{2}\mu v^2-\mu ve^2/C=\frac{1}{2}C^2/(\mu r^2)-\mu e^4/C^2=\frac{1}{2}e^2C^2/(\mu e^2 r^2)-e^2/r$ | $\frac{1}{2}\mu v^2-e^2/r=\frac{1}{2}\mu e^4/(n^2\hbar^2)-\mu e^4/(n^2\hbar^2)=-\frac{1}{2}\mu e^4/(n^2\hbar^2)=-R_H/n^2$ | A13 |
| Feedback to dE/dr=0 at $r_0$ | $-C^2/(\mu r^3)+e^2/r^2$ or $C^2/\mu=e^2 r_0$ | absent | A14 |
| Feedback to E (A13) | $E=\frac{1}{2}e^2 r_0/r^2-e^2/r=\frac{1}{2}(e^2/r_0)[(r_0/r)^2-2r_0/r]$ | absent | A15 |
| Feedback to $d^2E/dr^2=k$ | $k=3C^2/(\mu r^4)-2e^2/r^3=3e^2 r_0/r^4-2e^2/r^3= 2(e^2/r_0^3)(r_0/r)^3[1,5(r_0/r)-1]$ | absent | A16 |
| Classical r definition using n | $r=nr_0$ | absent, replaced by A12 or $r=n^2 r_B$ | A17 |
| Plugging (A17) in k (A16) | $k_n=k_1(1/n^3)(1,5/n-1)$; $k_1=e^2/r_0^3$ | absent | A18 |
| Plugging (A17) in E (A15) | $E=\frac{1}{2}(e^2/r_0)[1/n^2-2/n]$ | absent | A19 |
| Adding $E_0=\frac{1}{2}(e^2/r_0)$ to (A19) | $E'=E_0+\frac{1}{2}(e^2/r_0)[1/n^2-2/n]=E_0(1-1/n)^2$ | absent | A20 |
| Replacing 1/n by (1-1/n) | $E'=E_0[1-(1-1/n)]^2=E_0/n^2$ | see result A13 | A21 |
| Energy difference, terms $T_n$ | $T_n=E_0-E_0/n^2=E_0(1-1/n^2)$ | $T_n=R_H-R_H/n^2=R_H(1-1/n^2)$ | A22 |
| Identical T formulae | n defined classically in (A17) | n in Bohr quantum hypothesis (A9) | A23 |

[a] HO is the classical Harmonic Oscillator

Force constant equations (A16)-(A18) for periodic motion and vibrations in HOs, are absent in Bohr theory. A switch to complementary variable (A21) is a switch from (i) energy $V=-e^2/r$ in (A1) to energy difference $\Delta V=-e^2/r+e^2/r_0$ and (ii) of moment $p=C/r$ in (A10) to moment difference $\Delta p=C(1/r-1/r_0)$. Kinetic and potential energy differences give $\Delta E=\frac{1}{2}(e^2/r_0)[\frac{1}{2}(1-1/n)^2-(1-1/n)]=\frac{1}{2}(e^2/r_0)/n^2$ (A21).

The usefulness of complementary variable 1-1/n in (A21), usually not considered for H theory, is illustrated by respective 4th order fits (2 decimal version) of $E'_n$ in Table 2

1/n:         $E'_n=-4,37/n^4+5,55/n^3+109677,59/n^2-0,00/n+0,00$ cm$^{-1}$
(1-1/n):     $E'_n=-4,37(1-1/n)^4+11,91(1-1/n)^3+109668,05(1-1/n)^2-219354,37(1-1/n)+109678,77$ cm$^{-1}$

Reducing H size classically in (A17) without a quantum theory gives the same results as Bohr's quantum hypothesis for angular momentum (A9).